\documentclass[runningheads]{llncs}
\usepackage[T1]{fontenc}
\usepackage{amssymb}
\usepackage{amsmath}
\usepackage{flushend}
\usepackage{graphicx}
\usepackage{balance}
\usepackage{algorithm}
\usepackage{algpseudocode}
\usepackage{comment,fourier}
\usepackage{todonotes}
\usepackage{tikz}
\usetikzlibrary{matrix}
\usetikzlibrary{shapes.geometric}
\tikzstyle{every node}=[font=\footnotesize]
\tikzstyle{etichetta}=[]
\tikzstyle{square}=[draw,outer sep=0pt,inner sep=-1.3em,regular polygon,regular polygon sides=4,,minimum size=13.5mm]
\tikzstyle{square2}=[square,dotted,black!50]
\pgfmathsetmacro{\ra}{0.95}

\usetikzlibrary{decorations.pathreplacing,calc}
\newcommand{\tikzmark}[1]{\tikz[overlay,remember picture] \node (#1) {};}

\newcommand*{\AddNote}[4]{%
    \begin{tikzpicture}[overlay, remember picture]
        \draw [decoration={brace,amplitude=0.5em},decorate,ultra thick,black]
            ($(#3)!(#1.north)!($(#3)-(0,1)$)$) --  
            ($(#3)!(#2.south)!($(#3)-(0,1)$)$)
                node [align=center, text width=2.5cm, pos=0.5, anchor=west] {#4};
    \end{tikzpicture}
}

\begin{document}

\title{BinomialHash: A Constant Time, Minimal Memory Consistent Hashing Algorithm}
\titlerunning{BinomialHash}

\author{Massimo Coluzzi\inst{1}\orcidID{0000-0001-5601-7870} \and
Amos Brocco\inst{1}\orcidID{0000-0002-0262-2044} \and
Alessandro Antonucci\inst{1}\orcidID{0000-0001-7915-2768} \and
Tiziano Leidi\inst{1}\orcidID{0000-0002-6335-7977}}

\authorrunning{M. Coluzzi et al.}
\institute{Department of Innovative Technologies,\\
        University of Applied Sciences and Arts of Southern Switzerland,\\
        Lugano, Switzerland
\email{\{massimo.coluzzi,amos.brocco,alessandro.antonucci,tiziano.leidi\}@supsi.ch}}

\maketitle  

\begin{abstract}
Consistent hashing is a technique for distributing data across a network of nodes in a way that minimizes reorganization when nodes join or leave the network. It is extensively applied in modern distributed systems as a fundamental mechanism for routing and data placement. Similarly, distributed storage systems rely on consistent hashing for scalable and fault-tolerant data partitioning. This paper introduces \emph{BinomialHash}, a consistent hashing algorithm that executes in constant time and requires minimal memory. We provide a detailed explanation of the algorithm, present a pseudo-code implementation, and formally establish its strong theoretical guarantees. Finally, we compare its performance against state-of-the-art constant-time consistent hashing algorithms, demonstrating that our solution is both highly competitive and effective, while also validating the theoretical boundaries.

\end{abstract}

\section{Introduction}
One of the benefits of a distributed system is the possibility of scaling up and down the overall computational or storage capacity by redistributing the load across different nodes. This load can be represented by files in a distributed storage, records in a distributed database, or requests in a web service. Due to the heterogeneous nature of the data managed by different distributed systems, we can refer to each item as a \emph{data unit}. A significant issue faced during the development and deployment of a distributed system is to avoid overloading specific nodes during operation. It is crucial to distribute data units evenly among the nodes and maintain a balanced distribution when the number of nodes can change dynamically or in the event of failures. In this regard, consistent hashing algorithms are renowned for their ability to distribute the load uniformly across the system and reduce the amount of data units that need to be reassigned when the system is scaled up or down.

Consistent hashing algorithms are widely used in various contexts, including resource distribution across a cluster of nodes, data sharding, and peer-to-peer systems. In such applications, reliability and high availability are generally ensured, as in data storage applications, or the systems are inherently fault-tolerant and include failover capabilities, as in distributed computing or load balancing applications (where data units can be assigned to any available node). Therefore, the algorithms discussed in this paper assume a reliable environment in which random node failures do not occur. More specifically we don't allow changes to happen randomly, but we expect that nodes are added or removed in a controlled and scheduled manner. This assumption does not impose a limitation, as all the algorithms can be extended to handle arbitrary node removals as demonstrated by \emph{MementoHash}~\cite{Coluzzi2023MementoHashAS}.

\section{Related work}
Consistent hashing was first introduced by Karger et al.~\cite{karger1997consistent} in the context of distributed caching and peer-to-peer systems to minimize data reallocation when nodes join or leave a cluster. Since its inception, consistent hashing has been enhanced to address various challenges. Techniques such as \emph{Rendezvous hashing}~\cite{thaler1996rendezvous}, introduced by Thaler and Ravishankar, aim to provide similar benefits while simplifying implementation. Variants like \emph{Multi-probe}~\cite{appleton2015multi} and \emph{JumpHash}~\cite{lamping2014fast} consistent hashing further improve load balancing and lookup efficiency over the original algorithm. More recently, approaches such as \emph{AnchorHash}~\cite{mendelson2020anchorhash} and \emph{DxHash}~\cite{dong2021dxhash} have been proposed to handle dynamic scalability while ensuring strong load guarantees. Finally, \emph{MementoHash}~\cite{Coluzzi2023MementoHashAS} introduces efficient mechanisms for stateful key redistribution while minimizing operational overhead. Every algorithm is designed to address a specific use case, leading to significant variations in features and performance. A comprehensive survey and comparison of the most prominent consistent hashing algorithms published between 1996 and 2022 was published in 2023~\cite{coluzzi2022ch-survey}.

Recently, consistent hashing algorithms that operate in constant time and are memory-efficient have been published, all demonstrating high performance under the assumption that random failures do not occur. 
In December 2023, Eric Leu introduced a new algorithm named \emph{PowerCH}~\cite{leu2023fast}, which uses minimal constant memory and computes a key lookup in constant time. Subsequently, in February 2024, Masson and Lee published \emph{FlipHash}~\cite{masson2024fliphash}, very similar in performance to \emph{PowerCH}. Finally, Ertl published \emph{JumpBackHash}~\cite{Ertl_2024}, also constant in time and memory usage, which additionally avoids floating point operations resulting in faster lookup operations. In this paper, we present a constant-time solution that also avoids floating-point operations, achieving performance similar to \emph{JumpBackHash}. Our approach is built on a clear and intuitive conceptual model, based on binary trees, which facilitates a straightforward demonstration of the algorithm’s properties.

\section{Preliminaries}\label{preliminaries}
In this paper, we address the problem of distributing data units evenly among the nodes of a cluster. It is possible to identify every data unit with a unique key and every node with a numeric identifier named a bucket. Therefore, we address the problem of mapping $k$ keys to $n$ buckets. In this context, hashing algorithms, which are deterministic functions that take an arbitrary amount of data as input and produce a fixed-length output called \emph{hash value} or \emph{digest}, are commonly utilized. The digest is typically a number within a range larger than the number of buckets, so it must be remapped to the interval $[1, n]$ using modular arithmetic. Hashing algorithms are designed to distribute keys evenly among buckets, ensuring balanced load distribution across all buckets and facilitating efficient determination of the mapping between keys and buckets. However, when the number of buckets changes, a simplistic approach based on hash functions and modular arithmetic results in a substantial remapping of keys to different buckets. As a result, a distributed system undergoes extensive relocation of data units across nodes each time a node joins or leaves the cluster. To mitigate this issue, consistent hashing solutions have been developed. A hashing algorithm is \emph{consistent} when the following properties are satisfied:

\begin{itemize}
    \item[] \textbf{balance}: keys are evenly distributed among buckets. Given $k$ keys and $n$ buckets, ideally, $\frac{k}{n}$ keys are assigned to each bucket.
    \item[] \textbf{minimal disruption}: the same key is always mapped to the same bucket as long as such a bucket is available. When a bucket leaves the cluster, only the keys mapped to such a bucket will move to other buckets.
    \item[] \textbf{monotonicity}: when a new bucket is added, keys only move from existing buckets to the new one.
\end{itemize}

\begin{note}\label{note:uniformHashFunctions}
\textbf{(Uniform hash functions)} We assume the hash functions used within consistent hashing algorithms to produce a uniform distribution of the keys.
\end{note}

\subsection{Notation}
As stated in previous sections, we can map each node of a distributed system to an integer in the range $[1,n]$ called a \emph{bucket}. Therefore, we can represent a cluster as an array of buckets named \emph{b-array}.
The considered algorithms do not support failures of random nodes in the cluster. Nodes can join or leave the cluster only in a \emph{Last-In-First-Out} (LIFO) order. 
In the proposed algorithm, we utilize the mapping between a b-array and a binary tree, which is the basis for the algorithm's name. Below, we provide a brief summary of key concepts related to binary trees.

\begin{definition}[\textbf{Perfect binary tree}]
    A \emph{perfect binary tree}~\cite{cormen2022introduction} is a binary tree where all levels are completely filled. In a \emph{perfect binary tree} every node has exactly two children except for the nodes in the lowest level that have no children.
\end{definition}

\begin{definition}[\textbf{Complete binary tree}]
    A \emph{complete binary tree}~\cite{cormen2022introduction} is a special kind of binary tree where insertion takes place level by level and from left to right order at each level. All levels of the tree are fully filled except for the terminal one, which may exhibit partial occupancy (see, e.g., Fig.~\ref{fig: complete binary tree}).
\end{definition}

\begin{figure}
  \begin{minipage}[t]{.49\textwidth}\centering
\begin{tikzpicture} [x=.1\textwidth,level distance=8mm,
    every node/.style={circle,draw,inner sep=0pt,minimum size=4mm},
    level 1/.style={sibling distance=20mm},
    level 2/.style={sibling distance=10mm},
    level 3/.style={sibling distance=5mm}]
    
    \node (n1) {1}
    child {
        node (n2) {2}
        child {
            node (n4) {4}
            child{ node (n8) {8} }
            child{ node (n9) {9} }
        }
        child{
            node (n5) {5}
            child{ node (n10) {10} }
            child{ node (n11) {11} }
       }
    }
    child {
        node (n3) {3}
        child {
            node (n6) {6}
                child [loosely dotted,font=\itshape,text=gray]{
                    node [loosely dotted,font=\itshape,text=gray] (n12) {12}
                }
                child [loosely dotted,font=\itshape,text=gray] {
                    node [loosely dotted,font=\itshape,text=gray] (n13) {13}
                }
        }
        child {
            node (n7) {7}
                child [loosely dotted,font=\itshape,text=gray]{
                    node [loosely dotted,font=\itshape,text=gray] (n14) {14}
                }
                child [loosely dotted,font=\itshape,text=gray] {
                    node [loosely dotted,font=\itshape,text=gray] (n15) {15}
                }
        }    
    };

    \node[left=4 of n1] (l1) [align=left,font=\scriptsize,draw=none] {Level 1}
    child {
        node (l2)[align=left,font=\scriptsize,draw=none] {Level 2}
        edge from parent [draw=none]
        child {
            node (l3) [align=left,font=\scriptsize,draw=none]{Level 3}
            edge from parent [draw=none]
            child {
                node (l4) [align=left,font=\scriptsize,draw=none]{Level 4}
                edge from parent [draw=none]
            }
        }
    };
    
    \node[right=3 of n1] (r1) [draw=none] {}
    child {
        node (r2) [draw=none] {}
        edge from parent [draw=none]
        child {
            node (r3) [draw=none] {}
            edge from parent [draw=none]
            child {
                node (r4) [draw=none] {}
                edge from parent [draw=none]
            }
        }
    };

    \draw[black,dotted,thick] (l1) -- (n1) (n1) -- (r1);
    \draw[black,dotted,thick] (l2) -- (n2) (n2) -- (n3) (n3) -- (r2);
    \draw[black,dotted,thick] (l3) -- (n4) (n4) -- (n5) (n5) -- (n6) (n6) -- (n7)  -- (r3);
    \draw[black,dotted,thick] (l4) -- (n8) (n8) -- (n9) (n9) -- (n10) (n10) -- (n11) -- (n12) -- (n13) -- (n14) -- (n15) -- (r4);
\end{tikzpicture}
\caption{Complete binary tree (11 nodes example)}
\label{fig: complete binary tree}
  \end{minipage}
  \hfill
  \begin{minipage}[t]{.49\textwidth}\centering
 \begin{tikzpicture} [x=.1\textwidth,level distance=8mm,
     every node/.style={circle,draw,inner sep=0pt,minimum size=4mm},
     level 2/.style={sibling distance=20mm},
     level 3/.style={sibling distance=10mm},
     level 4/.style={sibling distance=5mm}]

    \node (n0) {0}
    child {
        node (n1) {1}
        child {
            node (n2) {2}
            child {
                node (n4) {4}
                child{ node (n8) {8} }
                child{ node (n9) {9} }
            }
            child{
                node (n5) {5}
                child{ node (n10) {10} }
                child [loosely dotted,font=\itshape,text=gray] { node [loosely dotted,font=\itshape,text=gray] (n11) {11} }
           }
        }
        child {
            node (n3) {3}
            child {
                node (n6) {6}
                child [loosely dotted,font=\itshape,text=gray]{
                    node [loosely dotted,font=\itshape,text=gray] (n12) {12}
                }
                child [loosely dotted,font=\itshape,text=gray] {
                    node [loosely dotted,font=\itshape,text=gray] (n13) {13}
                }
            }
            child {
                node (n7) {7}
                child [loosely dotted,font=\itshape,text=gray]{
                    node [loosely dotted,font=\itshape,text=gray] (n14) {14}
                }
                child [loosely dotted,font=\itshape,text=gray] {
                    node [loosely dotted,font=\itshape,text=gray] (n15) {15}
                }
            }
        }
    };

    \node[left=3.5 of n0] (l0) [align=left,font=\scriptsize,draw=none] {Level 0}
    child {
        node (l1)[align=left,font=\scriptsize,draw=none] {Level 1}
        edge from parent [draw=none]
        child {
            node (l2) [align=left,font=\scriptsize,draw=none]{Level 2}
            edge from parent [draw=none]
            child {
                node (l3) [align=left,font=\scriptsize,draw=none]{Level 3}
                edge from parent [draw=none]
                child {
                    node (l4) [align=left,font=\scriptsize,draw=none]{Level 4}
                    edge from parent [draw=none]
                }
            }
        }
    };
    
    \node[right=3.5 of n0] (r0) [draw=none] {}
    child {
        node (r1) [draw=none] {}
        edge from parent [draw=none]
        child {
            node (r2) [draw=none] {}
            edge from parent [draw=none]
            child {
                node (r3) [draw=none] {}
                edge from parent [draw=none]
                child {
                    node (r4) [draw=none] {}
                    edge from parent [draw=none]
                }
            }
        }
    };

    \draw[black,dotted,thick] (l0) -- (n0) (n0) -- (r0);
    \draw[black,dotted,thick] (l1) -- (n1) (n1) -- (r1);
    \draw[black,dotted,thick] (l2) -- (n2) (n2) -- (n3) (n3) -- (r2);
    \draw[black,dotted,thick] (l3) -- (n4) (n4) -- (n5) (n5) -- (n6) (n6) -- (n7) (n7) -- (r3);
    \draw[black,dotted,thick] (l4) -- (n8) (n8) -- (n9) (n9) -- (n10) (n10) -- (n11) (n11) -- (n12) (n12) -- (n13) (n13) -- (n14) (n14) -- (n15) (n15) -- (r4);
    
\end{tikzpicture}
\caption{Hanging complete binary tree (11 nodes example)}
\label{fig: hanging tree example 1}
  \end{minipage}
\end{figure}
\begin{figure}[ht!]
\end{figure}

\begin{proposition}
    A \emph{complete binary tree} can consistently represent a \emph{b-array}. In this arrangement, the initial element located at position $1$ of the \emph{b-array} corresponds to the \emph{root} node of the \emph{complete binary tree}. Subsequently, for any position $i$ within the \emph{b-array}, the left child is identified as the element at position $2i$, while the right child is associated with the element at $2i+1$. For example, in Fig.~\ref{fig: complete binary tree} the children of node $5$ are node $10$ and node $11$.
\end{proposition}

\begin{definition}[\textbf{Capacity}]
    Given a \emph{binary tree} with $l$ levels, we call \emph{capacity} of such a tree the number of nodes of the corresponding \emph{perfect binary tree} with $l$ levels. Similarly, we define the \emph{capacity} of a level as the number of nodes of the same level in a \emph{perfect binary tree}. For example, the binary tree in Fig.~\ref{fig: complete binary tree} has capacity $15$ and its Level $4$ has capacity $8$.
\end{definition}

\begin{note} \label{note: complete tree properties}
    \textbf{(Properties of a complete binary tree)}
    If we represent a \emph{b-array} as a \emph{complete binary tree}, for every level $l$ of the tree, we can notice the following properties:
    \begin{enumerate}
        \item the capacity of the level is $2^{l-1}$;
        \item the value of the leftmost node of the level is $2^{l-1}$;
        \item the value of the leftmost node of the level represents the capacity of the level.
    \end{enumerate}
    Moreover, we can notice that the capacity of a tree with $l$ levels is $2^l-1$.
\end{note}

As detailed in Section \ref{sec: explanation of the algorithm}, it is crucial to maintain consistent key distribution across all levels of the tree structure. Specifically, to adhere to the principles of \emph{monotonicity} and \emph{minimal disruption}. For example, it is essential to ensure that a key assigned to bucket $1$ in a four-level tree is also assigned to bucket $1$ in trees with fewer levels, such as those with three or two levels.

To achieve such uniform behavior, we will leverage the properties of binary arithmetic and bitwise operations. To do that, we introduce a new data structure called the \emph{hanging complete binary tree} (or simply a \emph{hanging tree}). This structure includes an additional level, denoted as level $0$, which contains a single node with only one child (Fig.~\ref{fig: hanging tree example 1}). Similarly, we define a \emph{perfect hanging tree} as a \emph{perfect binary tree} with an added root that becomes the parent of the original root. We can map our \emph{b-array} to this data structure by using $0$-based indexing, thereby including bucket $0$ in the set of assignable buckets. Consequently, valid buckets will be in the range $[0,n\!-\!1]$.

\begin{proposition} \label{prop: hanging tree capacity}
    The capacity of a \emph{hanging tree} with $l\!+\!1$ levels is $2^l$.
\end{proposition}

\begin{note} (\textbf{Dealing with the hanging tree}) \label{note:level 0 special case}
    It is pertinent to observe that the properties expounded in Note~\ref{note: complete tree properties} remain valid for the revised tree structure, starting from level $1$ onward. Consequently, we will treat level $0$ as a special case while maintaining unmodified operations across all subsequent levels.
\end{note}

\section{Explanation of \emph{BinomialHash}} \label{sec: explanation of the algorithm}

In this section we describe the \emph{lookup} function (detailed in Alg.~\ref{alg: lookup}) which takes a key and returns a bucket in the range $[0,n\!-\!1]$. The core idea of the algorithm is to leverage the mapping between the \emph{b-array} and the \emph{hanging tree}. As shown in Fig.~\ref{fig: hanging tree example 1}, we can assume every level of the tree to be fully filled, except for the lowest one.

\begin{definition}
    Given a \emph{b-array} of size $n$, we assume buckets to be in the range $[0,n\!-\!1]$. We call \emph{enclosing tree} the smallest \emph{perfect hanging tree} whose number of nodes $E$ is a power of $\ 2$ and satisfies $E\geq n$. Similarly, we call \emph{minor tree} the largest \emph{perfect hanging tree} whose number of nodes $M$ is a power of $\ 2$ and satisfies $M<n$.
\end{definition}

\begin{proposition}
 \label{prop:notation}
    The following properties are true for every cluster size $n > 1$:
    \begin{itemize}
        \item the height of the \emph{enclosing tree} is $l = \lceil log_2(n) \rceil$;
        \item the capacity of the \emph{enclosing tree} is $E = 2^l$;
        \item the capacity of the \emph{minor tree} is $M = 2^{l-1}$;
        \item $E = 2M$ and $M < n \leq E$.
    \end{itemize}
\end{proposition}

\subsection{Basic idea}
To understand the underlying logic of the algorithm we consider a cluster of size $11$ (i.e., buckets are in the range $[0,10]$). We can represent the corresponding \emph{b-array} as the \emph{hanging tree} in Fig.~\ref{fig: hanging tree example 1}. In this case, the \emph{enclosing tree} is the \emph{perfect hanging tree} comprising buckets in range $[0,15]$ ($E=16$) while the \emph{minor tree} is the \emph{perfect hanging tree} comprising buckets in range $[0,7]$ ($M=8$). The lowest level of the \emph{enclosing tree} is $4$ while the lowest level of the \emph{minor tree} is $3$. The \emph{lookup} function hashes the key to obtain a numerical digest and then executes the following process to return a valid bucket.:

\paragraph{Step 1} The digest is mapped against the \emph{enclosing tree}, obtaining a value $b$ within the range $[0, E\!-\!1]$. Valid buckets are those within the range $[0, n\!-\!1]$; for instance, in the given example, the valid range is $[0, 10]$.

\paragraph{Step 2} Depending on the output $b$ of the previous step, the algorithm will either return a valid bucket or perform another iteration; more specifically:

\begin{enumerate}
    \item if $b \in [0, M\!-\!1]$: The bucket belongs to the \emph{minor tree}, where all buckets are valid (e.g., range \([0, 7]\) in the example). In this case, we rehash the key against the \emph{minor tree} for consistency and terminate.
    \item if $b \in [M, n\!-\!1]$: The bucket falls within the valid range of the lowest level (e.g., range $[8, 10]$ in the example). In this case, the bucket is returned directly.
    \item if $b \in [n, E\!-\!1]$: Buckets in this range are invalid (e.g., range $[11, 15]$ in the example). In this case, we repeat the process from \emph{Step 1} using a digest of the key produced by a different hash function.
\end{enumerate}

The entire process is repeated up to a fixed number of iterations, denoted by $\omega$. If, after all iterations, the result is still an invalid bucket within the range $[n, E\!-\!1]$, the process is terminated by rehashing the key against the \emph{minor tree}, which always guarantees a valid bucket.

\subsection{The algorithm in detail}
In this section, we provide a detailed description of the algorithm, while Alg.~\ref{alg: lookup} presents its pseudocode implementation. The algorithm initially hashes the input key to obtain an integer value (line $2$) and performs a bitwise $\mathtt{AND}$ operation between the hash value and the capacity of the \emph{enclosing tree} minus one (line $4$). As stated in Prop.~\ref{prop: hanging tree capacity}, the capacity of the \emph{enclosing tree} is $E = 2^l$, where $l$ is the height of the tree. Therefore, $E\!-\!1$ represents a bit mask capable of mapping any value in the range $[0,E\!-\!1]$. To improve balance across valid nodes, a relocation procedure (\emph{relocateWithinLevel},
which will be detailed in Sec.~\ref{sec: relocation function}) is used to uniformly distribute keys within the level where the bucket resides.

\begin{algorithm}
  \caption{Lookup}
  \label{alg: lookup}
  \begin{algorithmic}[1]
    \Ensure{$0 \leq b < n$}
    \Function{LOOKUP}{$key$}

      \State $h^0 \gets h \gets hash(key)$
      \For{$i\gets 0; i < \omega; i\gets i+1$}
        
        \State $b \gets h^i\ \mathtt{AND}\ (E-1)$
        \State $c \gets relocateWithinLevel(b,h^i)$
        \If{$c < M$} \tikzmark{a_top}
          \State $d\gets h\ \mathtt{AND}\ (M-1)$ 
          \State \Return $relocateWithinLevel(d,h)$\tikzmark{right}
        \EndIf\tikzmark{a_bottom}
       
        \If{$c < n$}\tikzmark{b_top}
          \State \Return $c$
        \EndIf\tikzmark{b_bottom}

        \State $h^{i+1} \gets hash^{i+1}(key)$
        
      \EndFor
      
      \State $d\gets h\ \mathtt{AND}\ (M-1)$ \tikzmark{c_top}
      \State \Return $relocateWithinLevel(d,h)$
      \tikzmark{c_bottom}
      
    \EndFunction
  \end{algorithmic}
  \AddNote{a_top}{a_bottom}{right}{A}
  \AddNote{b_top}{b_bottom}{right}{B}
  \AddNote{c_top}{c_bottom}{right}{C}
\end{algorithm}

 If the outcome of the first step returns a value in the range $[0, M\!-\!1]$, we remap the original hash value $h$ against the \emph{minor tree} (block~$A$). It should be noted that, even though the obtained bucket is already in the range $[0, M\!-\!1]$, we must rehash against the \emph{minor tree} to ensure \emph{monotonicity} and \emph{minimal disruption} when adding or removing a node causes the tree to change its number of levels (i.e., transitioning from $n=2^p$ to $n=2^p+1$ or vice versa). If the outcome is in the range $[M, n\!-\!1]$, we return the obtained bucket (block~$B$). As shown in Section~\ref{sec: analysis}, the last action preserves \emph{monotonicity} and \emph{minimal disruption}. The loop is executed up to a fixed number of iterations $\omega$. If neither block $A$ nor block $B$ applies in any iteration, the final part of the algorithm (block $C$) uniformly distributes the remaining keys in the range $[0, M\!-\!1]$, where all buckets are valid.
 From Prop.~\ref{prop: hanging tree capacity}, we know that the capacity of a \emph{hanging tree} with height $l$ is $2^l$ and the capacity of the lowest level $l$ is $2^{l-1}$. Consequently, the capacity of the lowest level is half the capacity of the whole tree. Hence, we know that the lowest level of a \emph{hanging tree} can hold as many buckets as the rest of the tree. If we consider the example in Fig.~\ref{fig: congruent remapping}, for every key not properly mapped during the loop, the block $C$ of the algorithm applies a congruent remapping of the key from the bucket $b$ in range $[8,15]$ into a bucket $d$ in the range $[0,7]$ by using a bit mask (Alg.~\ref{alg: lookup}, line $15$).
Finally, the algorithm relocates bucket $d$ within its level to be consistent with line $5$.
This last relocation is essential to guarantee \emph{monotonicity} and \emph{minimal disruption} when the number of levels in the tree changes. For example, if we start with a cluster of size $9$, the buckets are in the range $[0,8]$. Therefore, the \emph{b-array} is mapped to a \emph{hanging tree} with $4$ levels and capacity $16$ as depicted in Fig.~\ref{fig: hanging tree example 2}. Bucket $8$ is the only valid bucket in level $4$, therefore removing it will cause the \emph{hanging tree} to lose the lowest level. The last relocation guarantees that keys mapped to buckets from $0$ to $7$ before the removal of bucket $8$ keep the same position, and only the keys previously mapped to bucket $8$ are distributed uniformly within the range $[0,7]$.

\begin{figure}[t!]
  \begin{minipage}[t]{.49\textwidth}\centering
\centering
\begin{tikzpicture} [x=.1\textwidth,level distance=8mm,
    every node/.style={circle,draw,inner sep=0pt,minimum size=4mm},
    level 2/.style={sibling distance=30mm},
    level 3/.style={sibling distance=16mm},
    level 4/.style={sibling distance=6mm}]
    \node (n0) {0}
    child {
        node (n1) {1}
        child {
            node (n2) {2}
            child {
                node (n4) {4}
                child{ node (n8) {8} }
                child{ node (n9) {9} }
            }
            child{
                node (n5) {5}
                child{ node (n10) {10} }
                child [loosely dotted,font=\itshape,text=gray]{ node [loosely dotted,font=\itshape,text=gray] (n11) {11} }
           }
        }
        child {
            node (n3) {3}
            child {
                node (n6) {6}
                child [loosely dotted,font=\itshape,text=gray]{
                    node [loosely dotted,font=\itshape,text=gray] (n12) {12}
                }
                child [loosely dotted,font=\itshape,text=gray]{
                    node [loosely dotted,font=\itshape,text=gray] (n13) {13}
                }
            }
            child {
                node (n7) {7}
                child [loosely dotted,font=\itshape,text=gray]{
                    node [loosely dotted,font=\itshape,text=gray] (n14) {14}
                }
                child [loosely dotted,font=\itshape,text=gray]{
                    node [loosely dotted,font=\itshape,text=gray] (n15) {15}
                }
            }
        }
    };

    \node[right=.3 of n7] (r7) [draw=none] {};

    \draw[black,->,dotted,thick] (n8) .. controls (n8|-n4) and (n8|-n2) .. (n0);
    \draw[black,->,dotted,thick] (n9) -- (n1);
    \draw[black,->,dotted,thick] (n10) -- (n2);
    \draw[black,->,dotted,thick] (n11) .. controls (n11|-n5) and (n1|-n3) .. (n3);
    \draw[black,->,dotted,thick] (n12) .. controls (n12|-n3) and (n10|-n3) .. (n4);
    \draw[black,->,dotted,thick] (n13) -- (n5);
    \draw[black,->,dotted,thick] (n14) -- (n6);
    \draw[black,->,dotted,thick] (n15) .. controls (r7|-n15) and (r7|-n7) .. (n7);
    
\end{tikzpicture}
\caption{Congruent remapping}
\label{fig: congruent remapping}
 \end{minipage}
  \hfill
  \begin{minipage}[t]{.49\textwidth}\centering
\centering
 \begin{tikzpicture} [x=.1\textwidth,level distance=8mm,
     every node/.style={circle,draw,inner sep=0pt,minimum size=4mm},
     level 2/.style={sibling distance=20mm},
     level 3/.style={sibling distance=10mm},
     level 4/.style={sibling distance=5mm}]

    \node (n0) {0}
    child {
        node (n1) {1}
        child {
            node (n2) {2}
            child {
                node (n4) {4}
                child{ node (n8) {8} }
                child [loosely dotted,font=\itshape,text=gray] { node [loosely dotted,font=\itshape,text=gray] (n9) {9} }
            }
            child{
                node (n5) {5}
                child [loosely dotted,font=\itshape,text=gray] { node [loosely dotted,font=\itshape,text=gray] (n10) {10} }
                child [loosely dotted,font=\itshape,text=gray] { node [loosely dotted,font=\itshape,text=gray] (n11) {11} }
           }
        }
        child {
            node (n3) {3}
            child {
                node (n6) {6}
                child [loosely dotted,font=\itshape,text=gray] {
                    node [loosely dotted] (n12) {12}
                }
                child [loosely dotted,font=\itshape,text=gray] {
                    node [loosely dotted,font=\itshape,text=gray] (n13) {13}
                }
            }
            child {
                node (n7) {7}
                child [loosely dotted,font=\itshape,text=gray] {
                    node [loosely dotted,font=\itshape,text=gray] (n14) {14}
                }
                child [loosely dotted,font=\itshape,text=gray] {
                    node [loosely dotted,font=\itshape,text=gray] (n15) {15}
                }
            }
        }
    };

    \node[left=3.5 of n0] (l0) [align=left,font=\scriptsize,draw=none] {Level 0}
    child {
        node (l1)[align=left,font=\scriptsize,draw=none] {Level 1}
        edge from parent [draw=none]
        child {
            node (l2) [align=left,font=\scriptsize,draw=none]{Level 2}
            edge from parent [draw=none]
            child {
                node (l3) [align=left,font=\scriptsize,draw=none]{Level 3}
                edge from parent [draw=none]
                child {
                    node (l4) [align=left,font=\scriptsize,draw=none]{Level 4}
                    edge from parent [draw=none]
                }
            }
        }
    };
    
    \node[right=3.5 of n0] (r0) [draw=none] {}
    child {
        node (r1) [draw=none] {}
        edge from parent [draw=none]
        child {
            node (r2) [draw=none] {}
            edge from parent [draw=none]
            child {
                node (r3) [draw=none] {}
                edge from parent [draw=none]
                child {
                    node (r4) [draw=none] {}
                    edge from parent [draw=none]
                }
            }
        }
    };

    \node[square,draw=none] (s8) at (n8) {};
    \draw[black] (s8.north west) -- (s8.south east);
    \draw[black] (s8.south west) -- (s8.north east);
    
    \draw[black,dotted,thick] (l0) -- (n0) (n0) -- (r0);
    \draw[black,dotted,thick] (l1) -- (n1) (n1) -- (r1);
    \draw[black,dotted,thick] (l2) -- (n2) (n2) -- (n3) (n3) -- (r2);
    \draw[black,dotted,thick] (l3) -- (n4) (n4) -- (n5) (n5) -- (n6) (n6) -- (n7) (n7) -- (r3);
    \draw[black] (l4.west) -- (r4);
    
\end{tikzpicture}
\caption{Hanging complete binary tree: 9 nodes with removal}
\label{fig: hanging tree example 2}
  \end{minipage}
\end{figure}
\begin{figure}[ht!]
\end{figure}

\subsection{Relocation function} \label{sec: relocation function}
Relying only on congruent remapping to move keys from invalid buckets (in range $[n,E\!-\!1]$) over the range $[0, n\!-\!1]$ can result in an unbalanced distribution. This behavior causes some buckets in the range $[0,M\!-\!1]$ to have twice as many keys as other buckets (i.e., the keys originally mapped to bucket $b\!-\!M$ plus the keys mapped to bucket $b$).  Given the example in Fig.~\ref{fig: congruent remapping}, all keys initially mapped to buckets in range $[11,15]$ will be moved only to buckets in range $[3,7]$. To avoid such imbalance issues, we randomly shuffle the keys so that all keys initially assigned to an invalid bucket are uniformly distributed within the range $[0, M\!-\!1]$. We achieve this by relocating the keys at every level of the tree (function \emph{relocateWithinLevel} in Alg.~\ref{alg: lookup}). This way, keys initially sent to bucket $b$ are uniformly distributed among all the buckets at the same level. 
Taking as example the tree depicted in Fig.~\ref{fig: hanging tree example 1}, the keys mapped to bucket $12$ by line $4$ of Alg.~\ref{alg: lookup} are uniformly distributed across all buckets in range $[8,15]$ by line $5$. To ensure \emph{monotonicity} and \emph{minimal disruption} every bucket, returned by the \emph{lookup} function, must be relocated within its respective level of the tree (see Alg.~\ref{alg: lookup}, lines $5$, $8$, and $16$).

\begin{algorithm}
  \caption{Relocate within level}
  \label{alg: relocate}
  \begin{algorithmic}[1]
    \Function{relocateWithinLevel}{$b$, $h$}
      \If{$b < 2$}
        \State \Return $b$
      \EndIf
      \State $d \gets highestOneBitIndex(b)$
      \State $f \gets 2^d-1$
      \State $r \gets hash(h,f)$
      \State $i \gets r\ \mathtt{AND}\ f$
      \State \Return $2^d + i$
    \EndFunction
  \end{algorithmic}
\end{algorithm}

The relocation function is implemented by leveraging binary arithmetic principles. As delineated in Note~\ref{note: complete tree properties}, when a \emph{b-array} is represented as a \emph{hanging tree}, for every level $l$, the value of the leftmost node of $l$ is $2^{l-1}$, and the capacity of level $l$ is also $2^{l-1}$. The binary representation of a positive integer value $x$ is articulated as follows:
\begin{equation*}
    x = x_0\cdot 2^0+x_1\cdot2^1+x_2\cdot2^2+\dots+x_{n-1}\cdot2^{n-}1+x_n\cdot2^n
\end{equation*}
where, $n\geq0$ and $\forall i, x_i\in\{0,1\}$. Consequently, it is evident that, given a specific bucket $b$, its depth within the tree is determined by the index of the highest one-bit within the binary representation of $b$, and correspondingly, the level of $b$ is determined by its depth incremented by one. For instance, considering the bucket $11$ in Fig.~\ref{fig: hanging tree example 1}, its binary representation is depicted as:
\begin{equation*}
    11 = 1\cdot 2^0+1\cdot 2^1+0\cdot 2^2+1\cdot 2^3.
\end{equation*}
Thus, the depth of bucket $b$ is $3$, positioning it within level $4$.
Algorithm~\ref{alg: relocate} provides a possible implementation of the relocation function.
According to Note~\ref{note:level 0 special case}, we treat Level $0$ as a special case. Moreover, since Level $1$ counts only one node, we do not need to perform any relocation. Therefore, if the bucket is $0$ or $1$, belonging to either Level $0$ or Level $1$, we return it unmodified (Alg.~\ref{alg: relocate}, lines $2$ and $3$).
Otherwise, we compute its depth $d$ by finding the index of the highest one-bit. We create a bit mask $f = 2^d-1$ that will map any number into a value $i$ in the range $[0,2^d-1]$ if used in combination with the bitwise $\mathtt{AND}$ operator. Finally, we return $2^d+i$ which represents the relocated position of $b$ within the same level.
\subsection{Intrinsic imbalance}
In a perfectly balanced system, we expect each node to receive $\frac{k}{n}$ keys.  
The operations performed inside the loop (see Alg.~\ref{alg: lookup}, lines $3$ to $14$) uniformly distribute keys across the range $[0, n\!-\!1]$. Block~$C$ further distributes the remaining keys uniformly across the range $[0, M\!-\!1]$, causing the buckets in the range $[0, M\!-\!1]$ to hold slightly more keys than those in the range $[M, n\!-\!1]$. The extent of this imbalance is inversely proportional to the number of loop iterations. We will demonstrate in Section~\ref{sec: analisys-balance} that the number of unbalanced keys is reduced by more than half with each iteration. Consequently, the fraction of unbalanced keys is less than $\frac{1}{2^\omega}$, where $\omega$ is the maximum number of loop iterations. For example, setting $\omega=6$ ensures that the imbalance is less than $1.6\%$.

\section{Analysis} \label{sec: analysis}
In this section, we formally prove the properties of the algorithm. Specifically, we demonstrate that \emph{BinomialHash} performs lookups in constant time with minimal memory usage while ensuring consistent hashing properties, namely: \emph{monotonicity}, \emph{minimal disruption}, and \emph{balance}.

\subsection{Time and space complexity}
We assume that the keys being looked up are bounded in size. Consequently, the underlying hash function executes in constant time. The internal loop runs up to a fixed number of iterations. It is well known that the highest one-bit of a fixed-size word (e.g., a 64-bit register) can be determined in constant time~\cite{knuth1997art}. Given these assumptions, it is evident that Algorithm~\ref{alg: lookup} runs in constant time. Furthermore, since the algorithm does not rely on large data structures and remains stateless (i.e., it does not store data between lookups), its space complexity is clearly minimal and constant.

\subsection{Monotonicity}
In this section, we will prove that \emph{monotonicity} holds for \emph{BinomialHash}. The property of \emph{monotonicity} states that when a new bucket is added, keys only move from existing buckets to the new one. We will prove this property by induction on the number of buckets $n$.
\begin{itemize}
    \item \emph{Base step} ($n=1$). The property trivially holds for $n=1$; keys can only move from the single existing bucket to the new one.
    \item \emph{Inductive hypothesis} ($n$). We assume the property holds for $n \geq 1$.
    \item \emph{Inductive step} ($M<n<E$). When a new node joins the cluster, keys initially mapped to buckets in the range $[0, n\!-\!1]$ during the first iteration of the loop will retain their positions, as the mapping depends solely on the key and $E$. A fraction $\frac{k}{E}$ of the keys will be assigned to bucket $n$ by block~$B$. The remaining keys will proceed to a new iteration of the loop, where the same conditions apply. Finally, block~$C$ maps only keys within the range $[0, M\!-\!1]$. Since neither $M$ nor $E$ change when $M < n < E$ and a new node joins the cluster, keys will either retain their current positions or be reassigned to the new bucket $n$.  
    \item \emph{Inductive step} ($n=E$). Finally, when $n=E$, all the available buckets in the \emph{enclosing tree} are valid. The keys are uniformly distributed among the $E$ valid buckets during the first iteration of the loop. Since all buckets are valid, block~$C$ is not executed. To add a new bucket, a new level must be added to the tree. Consequently, the current \emph{enclosing tree} becomes the new \emph{minor tree}, and the new \emph{enclosing tree} will have twice the size of the current one. The boundaries of the new tree are $M'=E$ and $E'=2E$, and the new size of the cluster is $n+1=E+1=M'+1$. In this new configuration, the first iteration of the loop distributes half of the keys in the range $[0,M'\!-\!1]$ and $\frac{k}{E'}$ keys into bucket $M'$. Subsequent iterations further populate the buckets in the range $[0,M']$, while block~$C$ distributes the remaining keys within the range $[0,M'\!-\!1]$. The results obtained from lines~$5$, $8$, and $16$ of the algorithm are congruent, meaning that line~$5$ applied to $E$ and lines~$8$ and $16$ applied to $M'$ distribute the keys in exactly the same way. Therefore, keys either retain their positions or move to the new bucket $M'$.
\end{itemize}

\subsection{Minimal disruption}
In this section, we will prove that \emph{minimal disruption} holds for \emph{BinomialHash}. The property of \emph{minimal disruption} states that when a bucket is removed, only the keys previously mapped to that bucket should move to a new location. As in the case of monotonicity, we will prove this property by induction on the number of buckets $n$. 
\begin{itemize}
    \item \emph{Base step} (n=2). The property trivially holds for $n=2$; when we remove a node, keys can only move from the removed bucket to the single remaining bucket.
    \item \emph{Inductive hypothesis} ($n-1$). We assume the property holds for $n\!-\!1 \geq 1$.
    \item \emph{Inductive step} ($M+1<n<E$). Given a cluster of size $n+1$, when bucket $n$ leaves the cluster, keys initially mapped to buckets in the range $[0, n\!-\!1]$ during the first iteration of the loop will retain their positions. Keys mapped to bucket $n$ during the first iteration will proceed to the second iteration, where they are either remapped to the range $[0, n\!-\!1]$ or passed to subsequent iterations. Finally, block~$C$ maps keys only within the range $[0,M\!-\!1]$. Therefore, keys previously mapped to buckets in the range $[0, n\!-\!1]$ retain their positions, while keys previously mapped to bucket $n$ are redistributed among the nodes in the range $[0,n\!-\!1]$.
    \item \emph{Inductive step} ($n=M+1$). When $n=M+1$, the lowest level of the tree contains only one valid node. The first iteration of the loop distributes half of the keys in the range $[0,M\!-\!1]$ and $\frac{k}{E}$ keys into bucket $M$. Subsequent iterations continue to populate the buckets in the range $[0,M]$, while block~$C$ distributes the remaining keys within the range $[0,M\!-\!1]$. When bucket $M$ is removed, the last level of the tree must also be removed. Consequently, the current \emph{minor tree} becomes the new \emph{enclosing tree}, and the new \emph{minor tree} will be half the size of the current one. The boundaries of the new tree are $E'=M$ and $M'=\frac{M}{2}$, resulting in a new cluster size of $n-1=M=E'$. In this configuration, the keys are uniformly distributed among the $E'$ valid buckets during the first iteration of the loop. Since all buckets are valid, block~$C$ is not executed. The results from lines~$5$, $8$, and $16$ of Alg.~\ref{alg: lookup} are consistent, meaning that line~$5$ applied to $E'$ and lines~$8$ and $16$ applied to $M$ distribute the keys in exactly the same way. Therefore, keys previously mapped to buckets in the range $[0,M\!-\!1]$ retain their positions, while keys previously mapped to bucket $M$ are redistributed among the nodes $0$ to $M\!-\!1$.

\end{itemize}

\subsection{Balance} \label{sec: analisys-balance}
\emph{BinomialHash} begins by hashing the key against the enclosing tree, thus returning a bucket $b \leq E-1$. If $b<n$ the algorithm stops. 
As we cope with uniform hashing functions (see Note~\ref{note:uniformHashFunctions}), this happens with probability $\frac{n}{E}$. If this is not the case, a rehashing of the input key against the enclosing tree is performed. This is done for a maximum number of attempts $\omega$ and, if this is not enough to get a valid bucket $b<n$, a hashing over the minor tree is eventually done, this giving, by construction, a bucket $b<M$ before the algorithm stops. The probability of the algorithm requiring all the $\omega$ attempts before the hashing over the minor tree is clearly $(\frac{E-n}{E})^{\omega}$. The complement of this probability corresponds to a situation where, after a variable number of hashing over the enclosing tree, we obtain a bucket $b<n$. This probability mass is uniformly distributed over $n$ buckets. As this is the only way to end with a bucket in one of the $n-M$ buckets of the lowest level, we have:
\begin{equation}\label{eq:prob}
P(M\leq b<n)=\frac{n-M}{n} \left[ 1-\left(\frac{E-n}{E}\right)^{\omega}\right]\,.
\end{equation}
Say that $k$ denote the total number of keys to be processed by \emph{BinomialHash}. Let us compute the expected number of keys for each one of the $n$ buckets. While a balanced procedure would gives the expected value $\frac{k}{n}$ on each bucket, \emph{BinomialHash} produces different expectations on the buckets of the lowest level and on the other ones. In the first case, the mass in Eq.~\eqref{eq:prob} is uniformly split over the $n-M$ buckets of the lowest level, thus giving the expected value:
\begin{equation}\label{eq:k}
K'=\frac{P(M\leq b<n)}{n-M} k\,.
\end{equation}
By similar considerations, for the buckets of the minor tree we have instead the expected value $K=\frac{1-P(M \leq b< n)}{M} k$. By simple algebra it is possible to notice that, as expected, the imbalance is towards the inner tree and hence $K'<\frac{k}{n}<K$. Let us therefore consider the gap between the two expectations normalized by the expected value for the perfectly balanced case, i.e.,
\begin{equation}\label{eq:delta}
\frac{K-K'}{\frac{k}{n}}
=\frac{1}{2^{\omega}} \left(1+\frac{n-M}{M} \right)
 \left(1-\frac{n-M}{M}\right)^{\omega}
\,,
\end{equation}
where the expression of the r.h.s., which is independent of $k$, can be obtained by simple algebra from Eq.~\eqref{eq:prob} and Eq.~\eqref{eq:k}.
If we regard Eq.~\eqref{eq:delta} as a function of $n\in [M,2M[$, it is easy to check that is a monotonically decreasing function, whose maximum has value
$2^{-\omega}$, achieved for $n=M$. This is an upper bound for the relative imbalance and its value decreases (exponentially) for an increasing number of iterations $\omega$.

Finally, we consider the variance, which in our setup takes form:
\begin{equation}\label{eq:var}
\sigma^2 =
\frac{M
\left(
\frac{k}{n}-K \right)^2 +
(n-M) \left(
K' -
\frac{k}{n} \right)^2}{n} \,.
\end{equation}
From Eq.~\eqref{eq:var}, by Eq.~\eqref{eq:k}, we get the following expression for the standard deviation:
\begin{equation}\label{eq:sigma}
\sigma(n,k)=\frac{k}{n} \sqrt{\frac{n-M}{M}
\left(\frac{2M-n}{2M}\right)^{\omega}}\,.
\end{equation}
If $k=q \cdot n$, i.e., we keep constant and equal to $q$ the expected number of keys per bucket in the balanced case, Eq.~\eqref{eq:sigma} has a single maximum achieved for $n = \frac{2+\omega}{1+\omega} M$ and equal to:
\begin{equation} \label{eq: sigma-max}
\sigma_{\mathrm{max}}= q \sqrt{\frac{1}{1+\omega}
\left(\frac{\omega}{2(1+\omega)}\right)^{\omega}}\,,
\end{equation}
that is a monotonically decreasing function of $\omega$ such that $\sigma_{\mathrm{max}}\simeq 0.045 q$ for $\omega=5$.

\section{Benchmarks}
To empirically validate the results presented in Section \ref{sec: analysis}, we conducted a benchmark comparing the performance of \emph{BinomalHash} and three other constant-time consistent hashing algorithms, namely \emph{PowerCH}~\cite{leu2023fast}, \emph{FlipHash}~\cite{masson2024fliphash}, and \emph{JumpBackHash}~\cite{Ertl_2024}. Our benchmark were conducted on Intel® Core™ i7-1065G7 CPU with 4 cores and 8 threads, and 32GB of main memory; all the algorithms were implemented in Java to provide fair and comparable implementations, and are publicly available on GitHub \cite{isinGitHub} together with the benchmarking tools. The considered metrics are \emph{lookup time} (i.e., the time required for resolving the bucket which stores a particular key), and \emph{balance} (i.e., the ability to spread the keys evenly among the buckets). Benchmarks related to memory usage are not reported because all the considered algorithms are practically stateless (they do not store data between lookups) and do not rely on large data structures.

\begin{figure}
  \begin{minipage}[t]{.49\textwidth}
    \includegraphics[width=\textwidth]{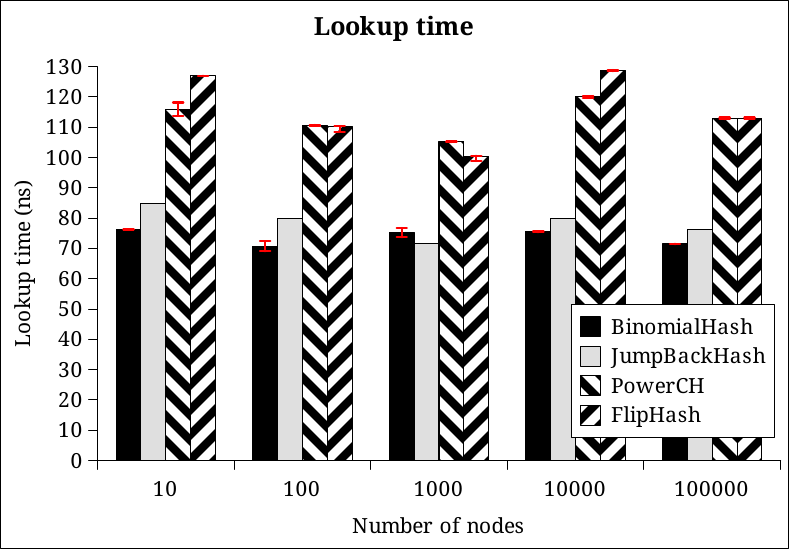}
    \caption{Lookup time}
    \label{fig:lookup_time}
  \end{minipage}
  \hfill
  \begin{minipage}[t]{.49\textwidth}
    \includegraphics[width=\textwidth]{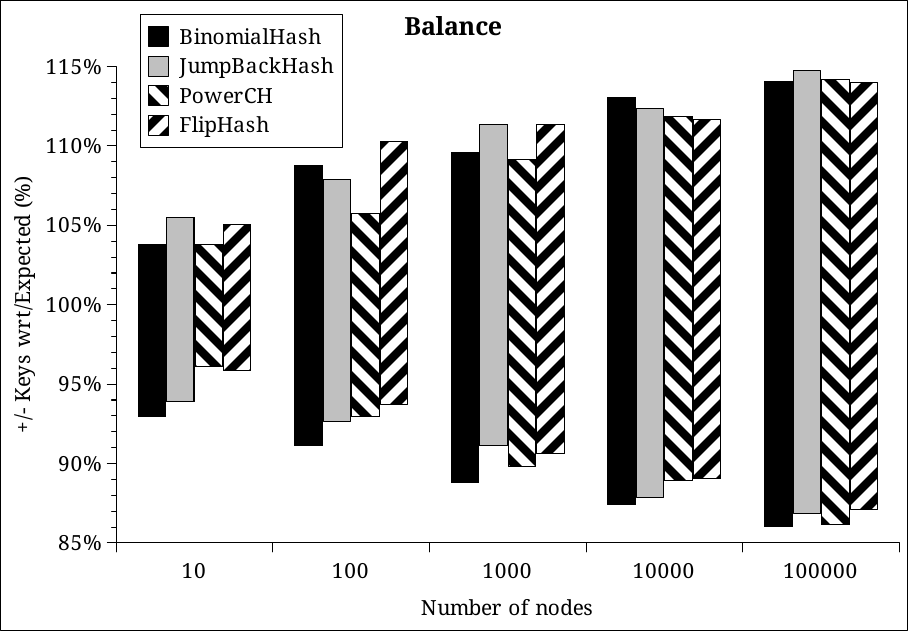}
    \caption{Relative difference (keys assigned to the least/most loaded node; mean=1000)}
    \label{fig:balance_relative}
  \end{minipage}
\end{figure}
\begin{figure}[ht!]
\end{figure}

\begin{figure}[H]
  \begin{minipage}[t]{.49\textwidth}
    \includegraphics[width=\textwidth]{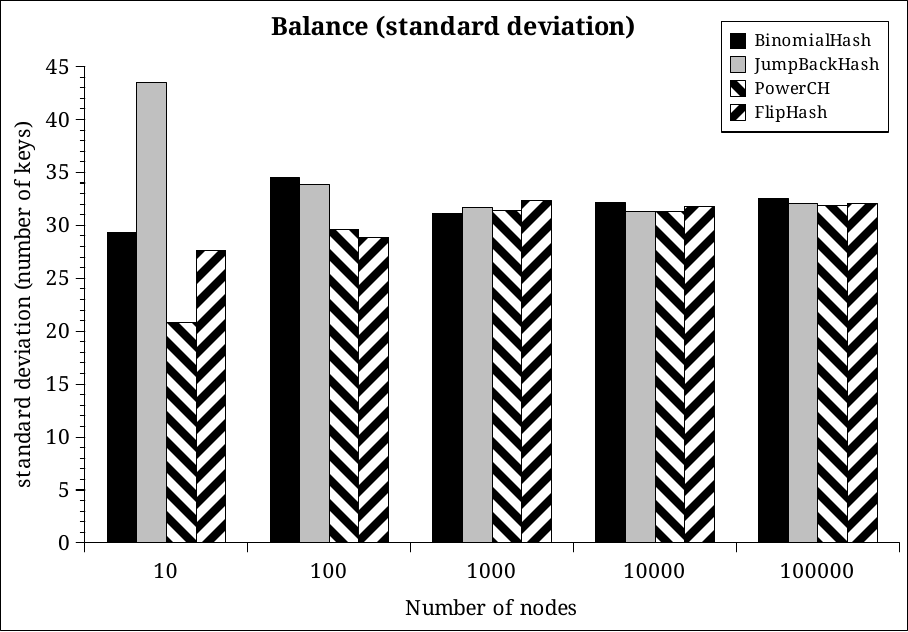}
    \caption{Standard deviation (mean=1000)}
    \label{fig:balance}
  \end{minipage}
  \hfill
  \begin{minipage}[t]{.49\textwidth}
    \includegraphics[width=\textwidth]{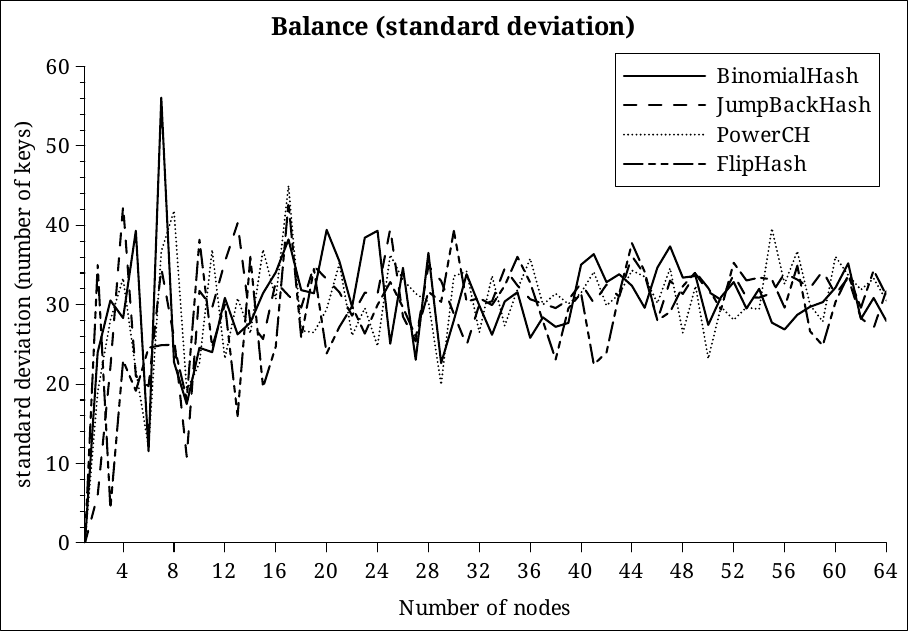}
    \caption{Standard deviation (mean=1000)}
    \label{fig:balance_64}
  \end{minipage}
\end{figure}

\paragraph{Lookup time.}
We performed lookups in clusters of varying scales, ranging from $10$ to $10^5$ nodes. Keys were sampled from a uniform distribution. As shown in Figure \ref{fig:lookup_time}, \emph{BinomialHash} is the best performer, alongside \emph{JumpBackHash}, while \emph{PowerCH} and \emph{FlipHash} are slightly slower. This difference could be attributed to the fact that both \emph{PowerCH} and \emph{FlipHash} use floating-point arithmetic, whereas \emph{BinomialHash} and \emph{JumpBackHash} rely solely on integer arithmetic.

\paragraph{Balance.}
We determined both the relative difference between the number of keys on the least loaded node and on the most loaded one (with respect to the average, namely $1000$ keys per node), and the standard deviation relative to the number of keys assigned to each node. As shown in Fig.~\ref{fig:balance_relative}, the difference is mild and no algorithm clearly dominates the others. Fig.~\ref{fig:balance} presents the results obtained from clusters of varying scales. Our experiments demonstrate that all algorithms perform similarly, achieving good balance across the network, with a relative standard deviation of less than $4\%$. This similarity is also evident in Fig.~\ref{fig:balance_64}, which details the standard deviation of the number of keys when scaling the cluster up to $64$ nodes. For \emph{BinomialHash}, these results are consistent with the bound of the maximum of the standard deviation derived in Eq.~\ref{eq: sigma-max}.

\section{Conclusions}
In this paper, we presented a novel constant-time consistent hashing algorithm named \emph{BinomialHash}, which offers an alternative approach to state-of-the-art algorithms such as \emph{PowerCH}~\cite{leu2023fast}, \emph{FlipHash}~\cite{masson2024fliphash}, and \emph{JumpBackHash}~\cite{Ertl_2024}. We provided implementation details and theoretical guarantees. Our solution enables lookups in constant time $O(1)$ while maintaining minimal memory usage. \emph{BinomialHash} is based on integer arithmetic, making it faster than \emph{PowerCH} and \emph{FlipHash}, which rely on floating-point arithmetic. Additionally, our solution is trivial to implement. We also provide benchmarks comparing \emph{BinomialHash} against the other three algorithms with respect to lookup time and balance. Like other constant-time consistent-hashing algorithms, \emph{BinomialHash} can be easily extended to handle arbitrary node removals and random failures by leveraging the procedure described in \emph{MementoHash}~\cite{Coluzzi2023MementoHashAS}.

\begin{credits}

\subsubsection{\discintname}
The authors have no competing interests to declare that are
relevant to the content of this article.
\end{credits}

\bibliographystyle{splncs04}
\bibliography{bibliography}

\begin{thebibliography}{10}
\providecommand{\url}[1]{\texttt{#1}}
\providecommand{\urlprefix}{URL }
\providecommand{\doi}[1]{https://doi.org/#1}

\bibitem{appleton2015multi}
Appleton, B., O'Reilly, M.: Multi-probe consistent hashing. arXiv preprint
  arXiv:1505.00062  (2015)

\bibitem{Coluzzi2023MementoHashAS}
Coluzzi, M., Brocco, A., Antonucci, A., Leidi, T.: Mementohash: A stateful,
  minimal memory, best performing consistent hash algorithm. IEEE/ACM Trans.
  Netw.  \textbf{32}(4),  3528–3543 (Apr 2024).
  \doi{10.1109/TNET.2024.3393476},
  \url{https://doi.org/10.1109/TNET.2024.3393476}

\bibitem{coluzzi2022ch-survey}
Coluzzi, M., Brocco, A., Contu, P., Leidi, T.: A survey and comparison of
  consistent hashing algorithms. In: 2023 IEEE International Symposium on
  Performance Analysis of Systems and Software (ISPASS). pp. 346--348 (2023).
  \doi{10.1109/ISPASS57527.2023.00048}

\bibitem{cormen2022introduction}
Cormen, T.H., Leiserson, C.E., Rivest, R.L., Stein, C.: Introduction to
  algorithms. MIT press (2022)

\bibitem{dong2021dxhash}
Dong, C., Wang, F.: Dxhash: A scalable consistent hash based on the
  pseudo-random sequence. arXiv preprint arXiv:2107.07930  (2021)

\bibitem{Ertl_2024}
Ertl, O.: <scp>jumpbackhash</scp>: Say goodbye to the modulo operation to
  distribute keys uniformly to buckets. Software: Practice and Experience  (Nov
  2024). \doi{10.1002/spe.3385}, \url{http://dx.doi.org/10.1002/spe.3385}

\bibitem{isinGitHub}
{Institute of Information Systems and Networking at SUPSI}:
  java-consistent-hashing-algorithms,
  \url{https://github.com/SUPSI-DTI-ISIN/java-consistent-hashing-algorithms}

\bibitem{karger1997consistent}
Karger, D., Lehman, E., Leighton, T., Panigrahy, R., Levine, M., Lewin, D.:
  Consistent hashing and random trees: Distributed caching protocols for
  relieving hot spots on the world wide web. In: Proceedings of the
  twenty-ninth annual ACM symposium on Theory of computing. pp. 654--663 (1997)

\bibitem{knuth1997art}
Knuth, D.E.: The Art of Computer Programming, Volume 1: Fundamental Algorithms.
  Addison-Wesley, Boston, MA, 3rd edn. (1997)

\bibitem{lamping2014fast}
Lamping, J., Veach, E.: A fast, minimal memory, consistent hash algorithm.
  arXiv preprint arXiv:1406.2294  (2014)

\bibitem{leu2023fast}
Leu, E.: Fast consistent hashing in constant time (2023)

\bibitem{masson2024fliphash}
Masson, C., Lee, H.K.: Fliphash: A constant-time consistent range-hashing
  algorithm. arXiv preprint arXiv:2402.17549  (2024)

\bibitem{mendelson2020anchorhash}
Mendelson, G., Vargaftik, S., Barabash, K., Lorenz, D.H., Keslassy, I., Orda,
  A.: Anchorhash: A scalable consistent hash. IEEE/ACM Transactions on
  Networking  \textbf{29}(2),  517--528 (2020)

\bibitem{thaler1996rendezvous}
Thaler, D., Ravishankar, C.V.: A name-based mapping scheme for rendezvous. In:
  Technical Report CSE-TR-316-96, University of Michigan. University of
  Michigan (1996)

\end{thebibliography}

\end{document}